\documentclass[aps,prl,twocolumn,superscriptaddress,10pt,english,nofootinbib,preprintnumbers]{revtex4-1}
\pdfoutput=1
\usepackage[T1]{fontenc}
\usepackage[utf8]{inputenc}
\usepackage{epsfig}
\usepackage{graphicx}
\usepackage{xcolor}
\usepackage{latexsym}
\usepackage{textcomp}
\usepackage{amssymb}
\usepackage[colorlinks=true,linkcolor=blue,citecolor=blue, urlcolor=blue]{hyperref} 
\usepackage{amsfonts,amsthm,amstext,amscd}
\usepackage{amsmath}
\usepackage{mathtools}

\def\trento{T\raisebox{-0.5ex}{R}ENTo}

\def\tr{\emph{Trajectum}\ }
\def\sf{$\sigma_\text{fluct}$}

\def\mpt{$\langle p_T \rangle$}

\def\trento{T\raisebox{-0.5ex}{R}ENTo\ }

\begin{document}
\title{Ultracentral heavy ion collisions, transverse momentum and the equation of state}
\author{Govert Nijs}
\affiliation{Theoretical Physics Department, CERN, CH-1211 Gen\`eve 23, Switzerland}
\author{Wilke van der Schee}
\affiliation{Theoretical Physics Department, CERN, CH-1211 Gen\`eve 23, Switzerland}
\affiliation{Institute for Theoretical Physics, Utrecht University, 3584 CC Utrecht, The Netherlands}
\affiliation{NIKHEF, Amsterdam, The Netherlands}
\begin{abstract}
Ultracentral heavy ion collisions due to their exceptionally large multiplicity probe an interesting regime of quark-gluon plasma where the size is (mostly) fixed and fluctuations in the initial condition dominate. Spurred by the recent measurement of the CMS collaboration we investigate the driving factors of the increase of transverse momentum, including a complete analysis of the influence of the QCD equation of state. Particularly interesting is the influence of the centrality selection as well as the initial energy deposition.
\end{abstract}

\preprint{CERN-TH-2023-234}

\maketitle

\noindent
{\bf Introduction -}
The hydrodynamic description of the quark-gluon plasma (QGP) formed in relativistic heavy ion collisions at Brookhaven National Laboratory and the Large Hadron Collider at CERN is one of the great successes of our understanding of many-body physics within Quantum Chromodynamics. This understanding relies on a relatively advanced interplay between an initial far-from-equilibrium stage, a viscous hydrodynamic stage, a switch to the hadronic phase and finally a relatively intricate global analysis of among else the anisotropies of the spectra of the detected particles \cite{Busza:2018rrf, Romatschke:2017ejr, Bernhard:2019bmu}. 

\begin{figure}[h]
\includegraphics[width=0.42\textwidth]{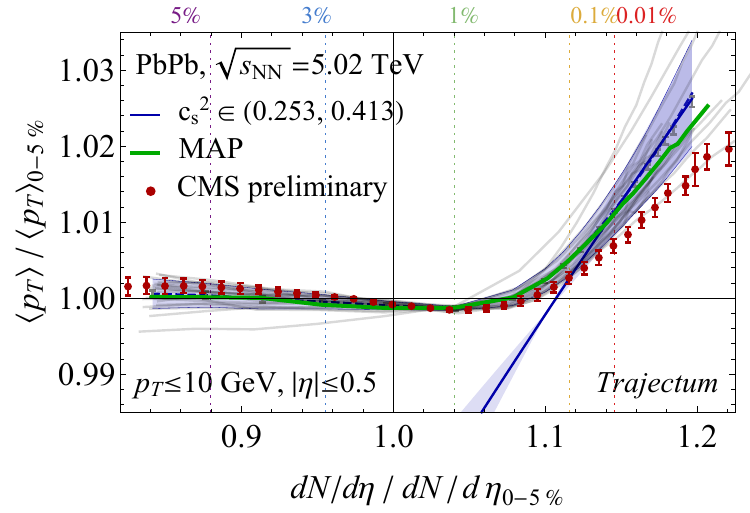}
\caption{\label{fig:cmsmainresult} We show the mean transverse momentum versus multiplicity for ultracentral events (up to the 0--0.001\% class) normalized by the central 0--5\% class for twenty likely parameter settings (grey)\@. The blue band is averaged from the grey lines (1$\sigma$) together with a fit to Eqn.~\eqref{eq:ptcs2}\@. The green line shows our most likely result (MAP)\@. CMS preliminary results are shown in red \cite{CMS:2023byumanual, CMS:2023byutalk}\@. The characteristic upward slope can among else be understood by the speed of sound at the initialization temperature (see text) \cite{Gardim:2019brr, Gardim:2019xjs}.}
\end{figure}

Recently, a novel way to look at the hydrodynamic QGP phase has been suggested by looking at the mean transverse momentum of the top 1\% highest multiplicity events as a function of that multiplicity \cite{Gardim:2019brr, Gardim:2019xjs} (see also \cite{Samanta:2023amp})\@. These head-on collisions have a (relatively) fixed transverse size and multiplicity fluctuations are mostly driven by initial entropy and temperature fluctuations. According to hydrodynamics the multiplicity is directly related to the entropy \cite{Muller:2005en} and the transverse momentum to the temperature  \cite{Gardim:2019xjs}\@. As such their relation has direct information on the equation of state (EoS) \cite{Gardim:2019brr}\@. Together with a later more refined hydrodynamic prediction from \tr \cite{Nijs:2021clz, Giacalone:2023cet} this has recently been measured by the CMS collaboration \cite{CMS:2023byumanual, CMS:2023byutalk} (see also results by ATLAS \cite{ATLAS:2023xpwmanual}), the result of which is shown in Fig.~\ref{fig:cmsmainresult}.

It is important that both the experimental and theoretical systematic uncertainties here are extremely small, often below 0.1\%\@. The reason is that Fig.~\ref{fig:cmsmainresult} shows a ratio with respect to `normal' central collisions in the \mbox{0--5\%} class and that in this ratio experimental or theoretical systematic uncertainties mostly cancel. The remaining non-trivial slope is conjectured to equal the speed of sound $c_s^2$ at the temperature at the start of hydrodynamics \cite{Gardim:2019brr}.

Given the importance of a possible direct extraction of the EoS this Letter will give a complete overview regarding this measurement, including its relation to the speed of sound. We will find that the EoS is only a contributing factor, with important contributions from both the initial state and the exact way how the centrality classes are determined.

{\bf Varying the EoS -}
In this work we will use the \tr framework, which consists of a generalized \trento initial stage \cite{Moreland:2014oya, Nijs:2023yab}, a viscous hydrodynamic stage and conversion to hadrons that are then evolved using SMASH \cite{SMASH:2016zqf,dmytro_oliinychenko_2020_3742965,0710.3820}\@. A complete description can be found in \cite{Nijs:2020roc, Nijs:2023yab, Giacalone:2023cet}, whereby we take the most recent parameter settings from \cite{Giacalone:2023cet}\@. In this particular work we use SMASH with only resonance decays without the collision term. 
In the appendix we verify that the collision term does not affect the results presented.

In order to test the conjecture that \cite{Gardim:2019brr}
\begin{equation}
    p_T \propto N_{\rm ch}^{c_s^2} \label{eq:ptcs2}
\end{equation}
the most obvious attempt is to vary the speed of sound and see how Eqn.~\eqref{eq:ptcs2} is affected. In this work we always label fits to Eqn.~\eqref{eq:ptcs2} by $c_s^2$ even though we will find that a one-to-one correspondence with the EoS is not always guaranteed.

Varying the EoS in heavy ion collisions was first done in \cite{Sangaline:2015isa}, where also a non-trivial posterior was obtained using a global analysis that agreed with results from the lattice. Here, this is done in a slightly different way. In \tr we follow \cite{Huovinen:2009yb, Bernhard:2018hnz}, which constructs a continuous hybrid EoS that interpolates between the hadron resonance gas (HRG) at low temperatures to the lattice result by HotQCD at high temperatures \cite{HotQCD:2014kol}. This is done in a somewhat intricate way, starting with the HotQCD 10-parameter fit of the pressure $P$ as function of the temperature $T$,
\begin{equation}\label{eq:eos}
  \frac{P}{T^4} = \frac{1}{2} \bigl( 1 + \tanh[c_t(t - t_0)] \bigr)
  \biggl( \frac{p_\text{id} + a_n/t + \cdots + d_n/t^4}{1 + a_d/t + \cdots + d_d/t^4} \biggr),
\end{equation}
where $t = T/T_c$, $T_c = 154$ MeV and $p_\text{id} = 95\pi^2/180 = \tfrac{\pi^2}{90}N_{\rm dof}$ is the weakly coupled limit for massless 3-flavor QCD\@. The ten parameters $t_0$, $c_t$ and $a_n$ till $d_d$ are given in \cite{HotQCD:2014kol}\@. Next, using thermodynamic identities the trace anomaly ($e - 3P$) is computed and smoothly connected with the HRG trace anomaly. Finally, from a starting temperature of 50 MeV the EoS is computed by integrating the trace anomaly \cite{Bernhard:2018hnz}\@. It is important that only for the (physical) parameters in \cite{HotQCD:2014kol} this gives back the same $P/T^4$ as in Eqn.~\eqref{eq:eos}, where in particular the parameter $p_{\rm id}$ is a measure for the number of degrees of freedom.

In the current work we vary the EoS in two ways. First, we vary both the parameters $p_{\rm id}$ and $a_n$ such that $P/T^4$ in the UV stays fixed. 
This is motivated by our prior knowledge that at high temperatures the EoS is well constrained by our knowledge of perturbative QCD. We note that this automatically limits large variations in the speed of sound, since the integral of the speed of sound determines the effective number of degrees of freedom, and is thereby quite unlike what is done in \cite{Sangaline:2015isa}\@. Secondly, more similar to \cite{Sangaline:2015isa}, we vary exclusively $p_{\rm id}$ to see the effect of the number of degrees of freedom on the speed of sound and the final hadronic observables.

\begin{figure}[t]
\includegraphics[width=0.49\textwidth]{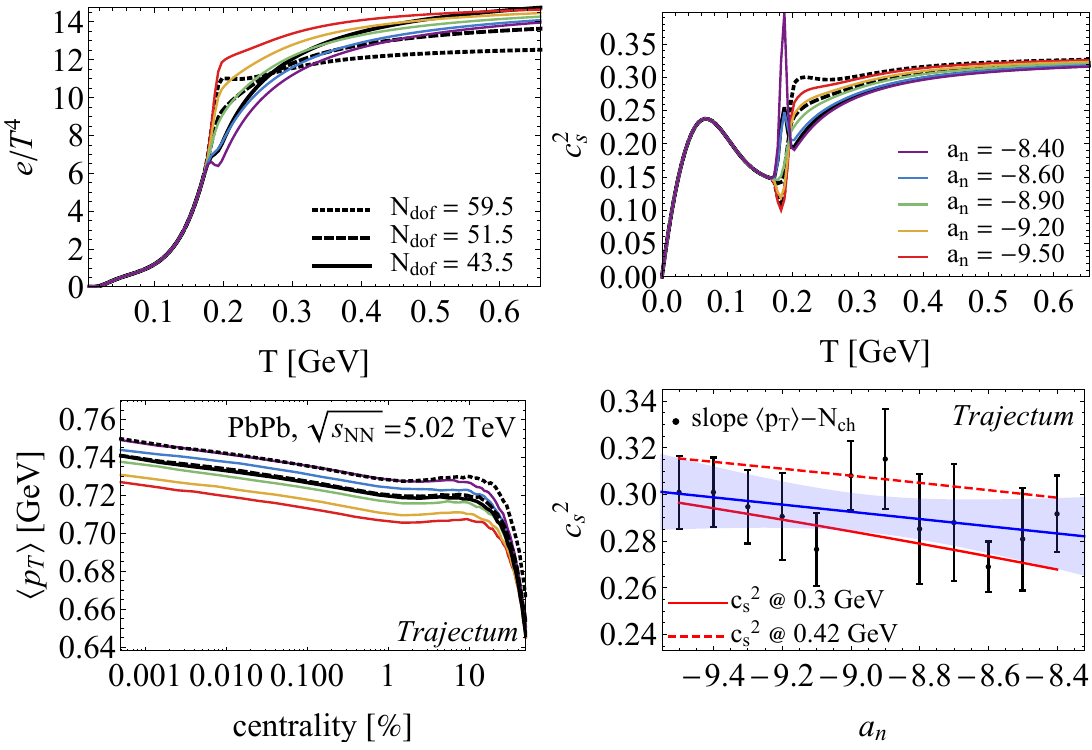}
\caption{\label{fig:varyeosall}(top) We show the energy density and speed of sound as a function of temperature for the variations as described in the text. (bottom) The varying EoS has a significant effect on \mpt{}, but less so on the slope as in Fig.~\ref{fig:cmsmainresult}\@. (bottom-right) Fitting the $c_s^2$ using Eqn.~\eqref{eq:ptcs2} we nevertheless see a decreasing trend that is consistent with the EoS trend (top) at temperatures 0.3 and $0.42\,$GeV (shown in red).}
\end{figure}

Fig.~\ref{fig:varyeosall} shows the resulting EoS for several parameter values (top), its \mpt{} as a function of centrality, and the fitted values using Eqn.~\eqref{eq:ptcs2} together with the actual values of $c_s^2$ at temperatures of $0.3\,$GeV and at the average ultracentral initialization temperature $\langle\langle T \rangle\rangle = 0.42\,$GeV\@. Here the first average is within an event, weighted by the energy density, and the second average is done in the 0--1\% centrality bin. It is clear that varying the EoS has a small but significant effect on the absolute value of \mpt{}, but only a small effect on the slope (bottom-left). Nevertheless, fitting $c_s^2$ to these variations we find within uncertainties that there is a decreasing trend that is consistent with the actual variations of the $c_s^2$ from the EoS\@. Ideally one would like larger variations of $c_s^2$, but as can be seen it is challenging to do so while keeping a causal $e/T^4$ that approaches the correct UV limit.

As shown in \cite{Giacalone:2023cet} the absolute change in \mpt{} and other observables can be compensated by variations in other parameters such as the viscosities. As such \cite{Giacalone:2023cet} did not obtain a posterior constraint on $a_n$, but some mild correlations exist between other parameters. It would be interesting to have more precise and direct knowledge of the EoS and Fig.~\ref{fig:varyeosall} suggests that the ultracentral \mpt{} can do this, but unfortunately only when the underlying precision is high.

\begin{figure*}[t!]
\includegraphics[width=0.85\textwidth]{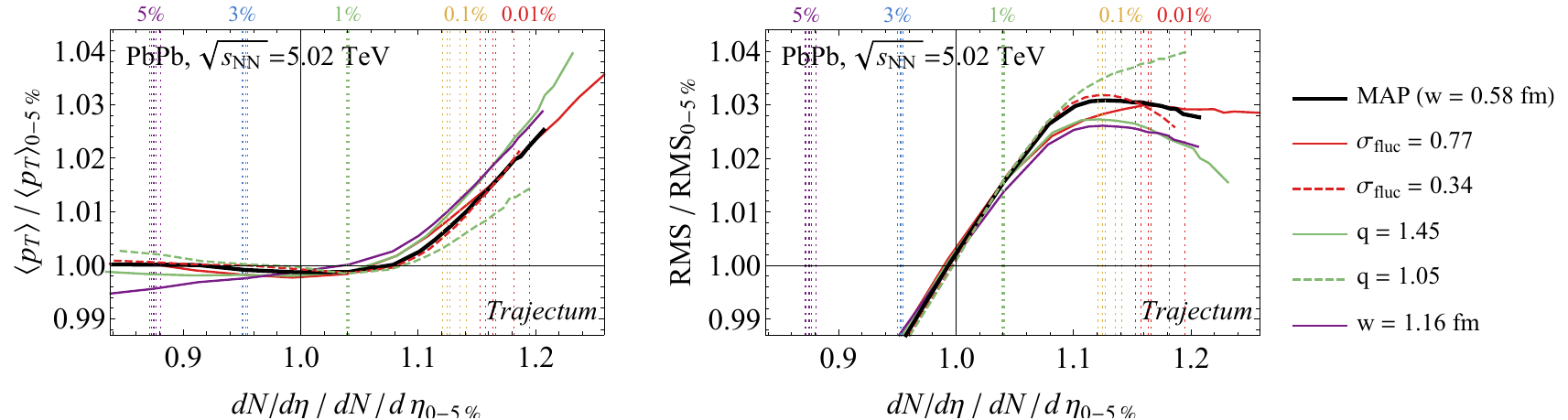}
\caption{\label{fig:varyparsmeanptratio}(left) As in Fig.~\ref{fig:cmsmainresult} we show the ultracentral \mpt{} for variations of the nucleon width $w$ (purple), the initial state fluctuations by \sf{} (red) and the \trento parameter $q$ (green)\@. The lines extend to the 0.001\% centrality bin, so that it is clear that more fluctuations (red) lead to a higher ultracentral multiplicity and \mpt{}, but do not change the slope. On the other hand and $q$ affects the slope significantly, which can be explained by the fact that ultracentral collisions in the case of large $q$ select a smaller initial QGP as measured by the RMS of the energy density (right).}
\end{figure*}

\begin{figure*}
\includegraphics[width=0.85\textwidth]{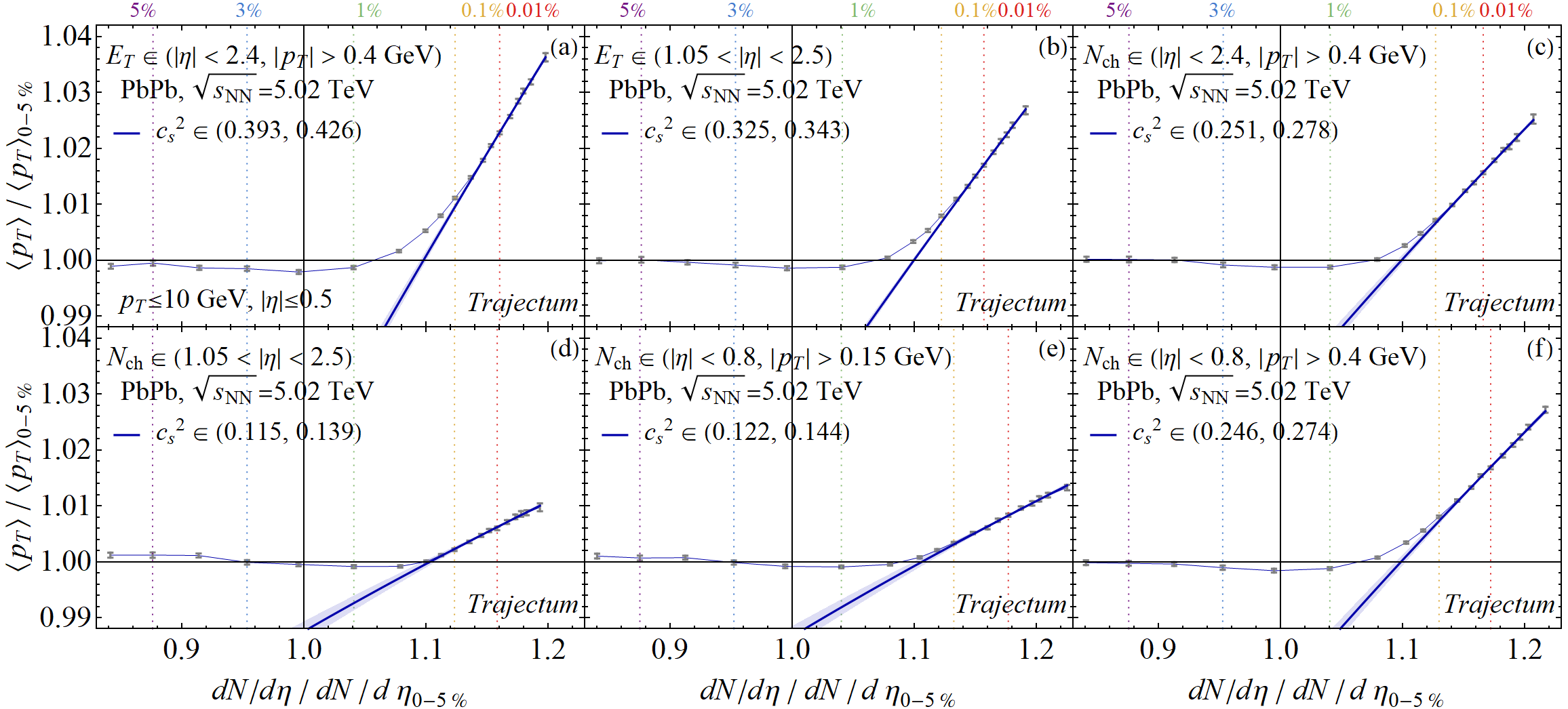}
\caption{\label{fig:varycentrselection} One of the central results of this Letter is that the slope of the ultracentral \mpt{} depends strongly on the centrality definition. Perhaps unsurprisingly, selecting centrality using a calorimeter (upper-left and upper-middle) causes a selection effect that favors a larger slope towards ultracentral collisions. This effect is stronger when defining centrality in the same region as the \mpt{} due to auto-correlation (upper-left)\@. Similar effects are present when using charged multiplicity, whereby especially a $p_T$ cut on multiplicity favors higher \mpt{} (right panels).}
\end{figure*}

{\bf Initial state and centrality definitions -}
Apart from the EoS we varied several other aspects that include the shear viscosity, bulk viscosity, nucleon size, the hydro starting time, initial energy fluctuations and the \trento{} initial state parameters $p$ and $q$\@. The latter parameters are used for a phenomenological description of the initial transverse energy density $e(x_\perp)$ as \cite{Moreland:2014oya,Nijs:2023yab}
\begin{equation}
\label{eq:e}
e(x_\perp) \propto \left(\frac{\mathcal{T}_L(x_\perp)^p + \mathcal{T}_R(x_\perp)^p}{2}\right)^{q/p},    
\end{equation}
in terms of the participant thickness functions of the left- and right-moving nuclei $\mathcal{T}_{L,R}$\@. The mean of $\mathcal{T}_{L,R}$ is proportional to the nucleon density of the nuclei and its fluctuations are given by a gamma distribution with parameter \sf{}\@.

Fig.~\ref{fig:varyparsmeanptratio} shows the analogue of Fig.~\ref{fig:cmsmainresult} while varying the nucleon width $w$, $q$ and \sf{}, which are the parameters that most directly influence the ultracentral ratios as shown. As anticipated in \cite{Gardim:2019brr} increasing \sf{} leads to a higher ultracentral \mpt{} and multiplicity (all curves stop at the class 0--0.001\%), but the slope is unaffected. Perhaps surprisingly, the $q$ parameter defined above affects the slope most significantly. The reason is that a large $q$ in combination with the ultracentral selection favors a small initial droplet (Fig.~\ref{fig:varyparsmeanptratio}, right). This can be understood, as a large $q$ can mimic binary scaling energy deposition, which as opposed to participant energy deposition is most effective in depositing entropy when all nucleons are in a small area. A constant area is a crucial assumption in \cite{Gardim:2019brr}, as a smaller area automatically implies a stronger radial flow and higher \mpt{}\@. We find that a large nucleon width changes the shape of the curve, but does not change the slope significantly.

Another crucial aspect of the ultracentral \mpt{} is the (experimental) centrality selection (see also \cite{Nijs:2021clz, Samanta:2023amp, Samanta:2023kfk})\@. Traditionally centrality has been regarded as a measure for impact parameter, but the more modern view is event activity. Event activity can however be estimated in several ways, each looking at activity in different areas of phase space. The two most standard estimators are forward multiplicity (for ALICE V0 this is given by V0A ($2.8\leq\eta\leq5.1$) and V0C ($-3.7\leq\eta\leq-1.7$) or forward energy as measured by a calorimeter. Alternatively it is possible to use mid-rapidity activity (see e.g.~\cite{ATLAS:2022dov}), which however has the feature that centrality often self-correlates with the observable measured. Lastly, while experimentally challenging, a lack of detected spectators in the ZDC detector can also signal a more central collision. Within \tr we can mimic all these strategies and hence make an apples-to-apples comparison with the measurement.\footnote{For mid-rapidity versus forward centrality selection it is important that the hydrodynamics in this work is boost invariant. The difference between the two hence only includes thermodynamic fluctuations at freeze-out, but ignores fluctuations in the initial condition. This hence serves as a lower bound on the difference in the selection.}

Fig.~\ref{fig:varycentrselection} shows six different centrality definitions, varying between calorimetric (transverse energy) or charged multiplicity and varying rapidity from the forward region (avoiding auto-correlation with the \mpt{} measurement) to mid-rapidity. For the charged multiplicity it is important that detectors especially at mid-rapidity often are limited in $p_T$ range, which we hence also varied. The standard \tr centrality selection is done with charged multiplicity at mid-rapidity ($|\eta|<0.8$ and $p_T>0.4\,$GeV), which is used in all other plots.\footnote{Experiments like ALICE almost always use forward centrality to avoid auto-correlation; in \tr this is however in general often computationally expensive, as a larger rapidity range needs to be simulated.} Clearly, selecting centrality calorimetrically favors events with a larger \mpt{} and hence has a larger slope going towards the most ultracentral events. Importantly, defining centrality with charged multiplicity of relatively high $p_T$ particles (e.g. $p_T>0.4\,$GeV) has a similar effect of favoring high \mpt{} events. It is for this reason that our `standard' centrality selection works relatively well in Fig.~\ref{fig:cmsmainresult} even though CMS used a forward calorimetric centrality definition. Lastly, as expected especially for the calorimetric definitions the auto-correlated definition (e.g.~defining centrality at mid-rapidity just like \mpt{}) has a stronger effect on \mpt{}\@.

While perhaps less surprising than the centrality cuts, it is clear that the ultracentral \mpt{} slope also depends on the $p_T$ cut of \mpt{} itself. This is shown in the Appendix, but here we point out that an even more accurate theory-experiment comparison can be made by focusing on charged particles with $p_T>0.3\,$GeV, since CMS relies on extrapolation for particles with lower $p_T$\@. The fact that the agreement is slightly different from Fig.~\ref{fig:cmsmainresult} indicates that the CMS extrapolation differs from the \tr spectrum. In the Appendix we also made a comparison with the ATLAS result using $p_T>0.5\,$GeV.

\begin{figure}
\includegraphics[width=0.37\textwidth]{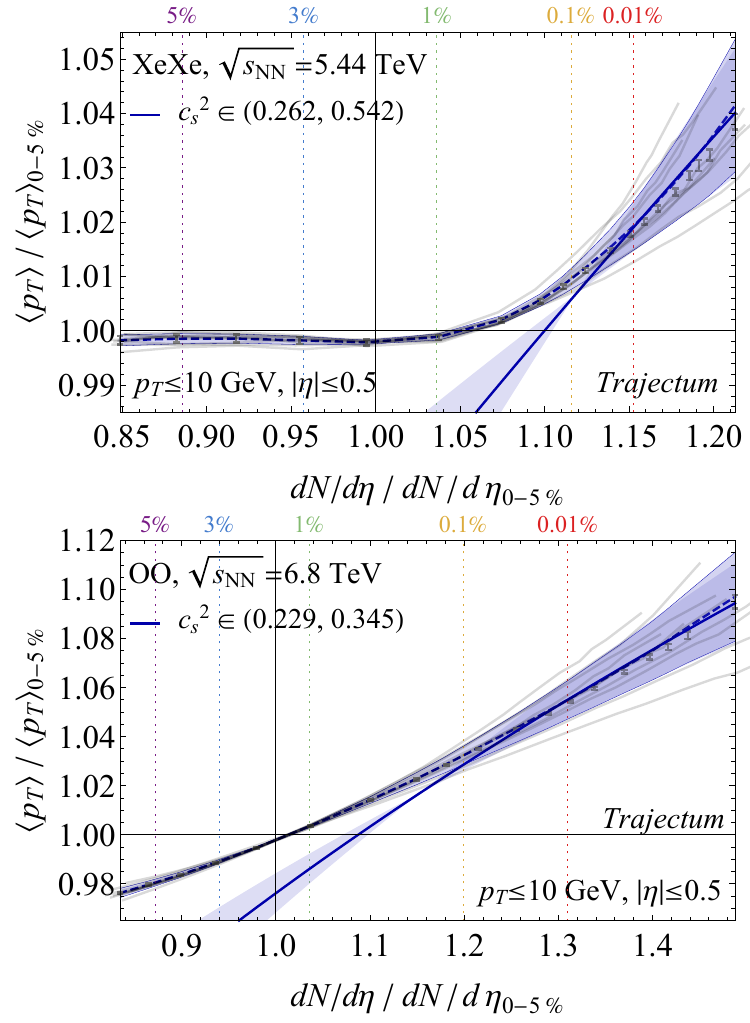}
\caption{\label{fig:xeandoxygen}We show \tr predictions for xenon (top) and oxygen (bottom) collisions at LHC energies analogous to Fig.~\ref{fig:cmsmainresult}.}
\end{figure}

{\bf Discussion -} 
One of the curious features of Fig.~\ref{fig:cmsmainresult} is the local minimum that arises around 1\% centrality. As far as we are aware there is no straightforward physical interpretation since all initial quantities such as average entropy density, average temperature, impact parameters, size of the QGP are monotonic around 1\% (see Appendix)\@. Nevertheless, around 1\% a qualitative change occurs where the impact parameter stops varying. 

At first the results of this work may seem disappointing. After all, the ultracentral \mpt{} may not be a direct measurement of the speed of sound. We, however, think this measurement may potentially be even more exciting. Hydrodynamics and the EoS are a crucial aspect of the increase in the \mpt{}, but several other factors contribute, including subtle size variations (depending on parameters, such as $q$ discussed before) as well as experimental considerations that includes the centrality definition. The fact that the final measurement with its precision due to experimental and theoretical uncertainty cancellations was predicted shows the importance and success of understanding all stages of heavy ion collisions at high precision and in particular highlights a novel possibility to constrain the initial energy deposition.

In the future it would be especially interesting to study the ultracentral \mpt{} in other systems, such as xenon (see also \cite{ATLAS:2023xpwmanual}) and oxygen. Predictions from \tr are provided in Fig.~\ref{fig:xeandoxygen}, where it can be seen that the xenon uncertainty is considerably wider (presumably due to its non-trivial shape \cite{Bally:2021qys, Bally:2022vgo})\@. Oxygen collisions fluctuate more due to their size, but the slope is relatively robust and consistent with PbPb\@. Seeing such a slope in Oxygen collisions at the LHC would be a unique signature of hydrodynamics in small droplets of QGP.

{\bf Acknowledgements -} We thank Cesar Bernardes, Somadutta Bhatta, Giuliano Giacalone, Wei Li, Jean-Yves Ollitrault and Jurgen Schukraft for interesting discussions.

\bibliographystyle{apsrev4-1}
\bibliography{letter, manual}
\clearpage
\newpage

\section{Appendix}

\begin{figure*}[h!]
\includegraphics[width=0.93\textwidth]{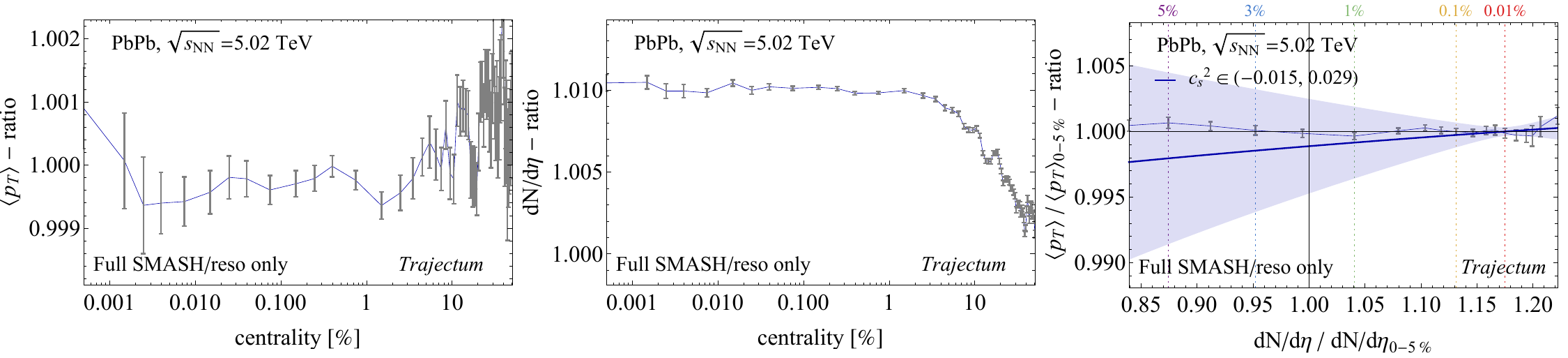}
\caption{\label{fig:withsmash}We show for the \mpt{} (left) and charged multiplicity (middle) as a function of centrality the ratio of the full SMASH result with respect to the approximation of doing resonance decays only. Within statistical uncertainties the ratio does not depend strongly on centrality in the ultracentral regime, which is why in this work we used SMASH in resonance decay mode only (see right figure for the full ratio).}
\end{figure*}

\subsection{Full SMASH versus resonance only}
In this work we showed results using SMASH with resonance decays only, as opposed to a full hadronic cascade. The obvious advantage is that in this way the computation is up to twenty times faster since a single hydro event can now easily be used to generate 50 hadronic events without increasing the computational time too much. Especially with central collisions and a large rapidity range this would be prohibitively expensive using the full SMASH code.

In this short section about Fig.~\ref{fig:withsmash} we verify however that due to the ratios taken this approximation does not affect any of our results (for this we ran a single run with full SMASH with about 5M events)\@. The reason is relatively clear, since in the ratio with the 0--5\% class many of the variations cancel (Fig.~\ref{fig:withsmash} right)\@. Nevertheless, even without such cancellation it can be seen that the charged \mpt{} within uncertainties is almost unaffected by the hadronic cascade (Fig.~\ref{fig:withsmash} left)\@. The multiplicity with the full cascade is about 1\% higher (presumably due to proton-antiproton annihilation into five pions \cite{Garcia-Montero:2021haa}), but the centrality dependence of this effect is very mild. We note that for more complicated observables (such as proton elliptic flow) this can be more complicated, which is one more reason why the ultracentral \mpt{} is such an insightful observable.

\subsection{Geometry}

\begin{figure*}
\includegraphics[width=0.93\textwidth]{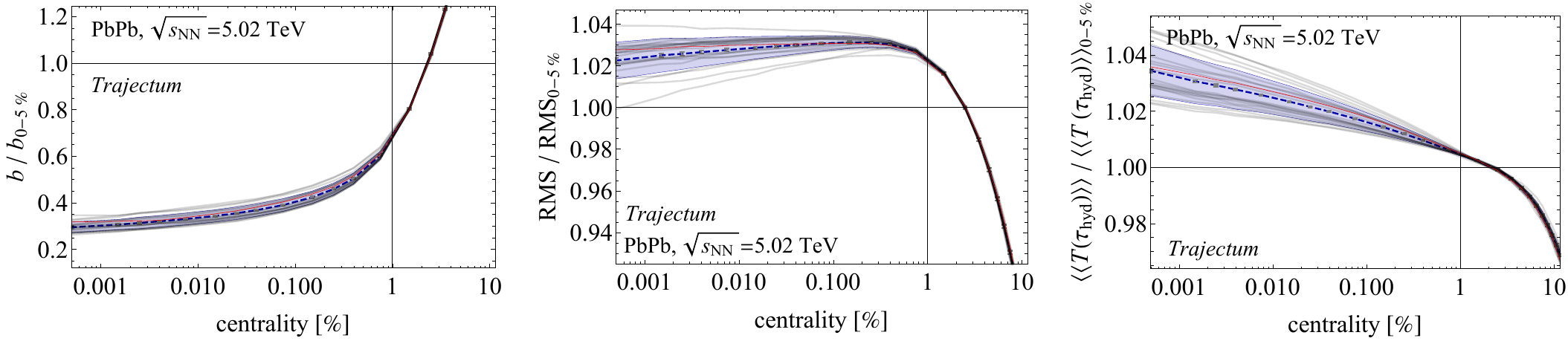}
\caption{\label{fig:geometryucratio}We show the impact parameter $b$, the RMS size of the initial QGP (middle) and the initial average temperature at the hydrodynamic initialization (right) as a function of centrality. Note that as in Fig.~\ref{fig:cmsmainresult} the curves are a variation of all parameters, including $\tau_{\rm hyd}$\@. As such the temperature will vary significantly among the parameters, but much less so when taking the ratio with respect to the 0--5\% class.}
\end{figure*}

For completeness we show the impact parameter $b$, the size of the initial droplet RMS and the initial temperature as a function of centrality and normalized to the 0--5\% class in Fig.~\ref{fig:geometryucratio}\@. Consistent with \cite{Gardim:2019brr} we see that towards ultracentral events the impact parameter becomes almost zero, the size becomes approximately constant and by construction (we select high multiplicity or high entropy states) the temperature keeps on increasing. Nevertheless, we see an indication that the size may be decreasing towards the 0.001\% class, which is related to the $q$ parameter as explained in the main text. If $q$ is sizeable then it is beneficial for a larger entropy to have participating nucleons distributed in a smaller space. Keeping everything fixed a smaller droplet leads to a higher \mpt{}\@. This can be understood both from a stronger radial flow (due to the larger gradients), or by noting that at fixed total entropy the smaller droplet has a larger temperature that in turn leads to a higher \mpt{} \cite{Gardim:2019xjs}.

\subsection{Transverse momentum cut variations}

\begin{figure*}
\includegraphics[width=0.85\textwidth]{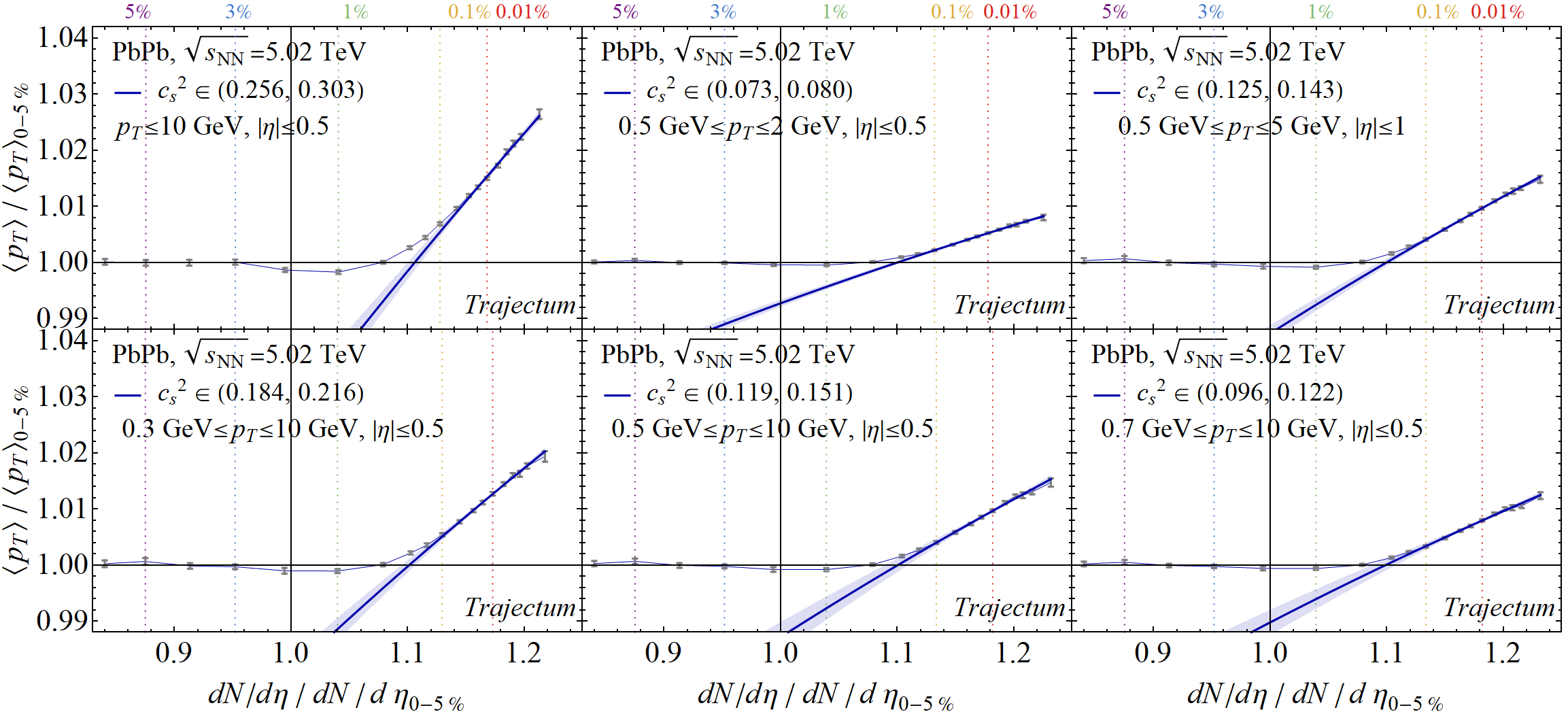}
\caption{\label{fig:varyptcut} In the main text we always show the inclusive \mpt{} (top-left, \tr does not produce hadrons with $p_T$ larger than $10\,$GeV)\@. In the bottom row we vary the lower $p_T$ cut, whereas the upper row also varies the upper cut-off.}
\end{figure*}

In Fig.~\ref{fig:varyptcut} we vary the cuts used to obtain both \mpt{} and the charged multiplicity for our MAP setting. It can be seen that the inclusive result in the main text (top-left) gives rise to the largest slope. Comparing the various cuts it can be seen that both tightening the lower or upper cut reduce the slope. This is unsurprising, since the larger \mpt{} is a result of shifting particles from lower to higher $p_T$ and as such a smaller interval will only capture part of this effect. While Eqn.~\eqref{eq:ptcs2} is conjectured for the inclusive \mpt{}, it is important that in experiment only particles above a minimum $p_T$ cut can be (accurately) measured. For ALICE this is typically $0.2\,$GeV and for CMS/ATLAS around $0.3\,$GeV\@. The lower range hence relies on theoretical modelling and a more direct comparison would compare for e.g.~the $p_T>0.3\,$GeV range. Secondly, mostly due to viscous corrections at freeze-out and hard QCD processes there is a considerable theoretical uncertainty for high-$p_T$ particles; for that reason it could be beneficial to e.g.~restrict to $p_T<2.0\,$GeV.

\begin{figure*}
\includegraphics[width=0.45\textwidth]{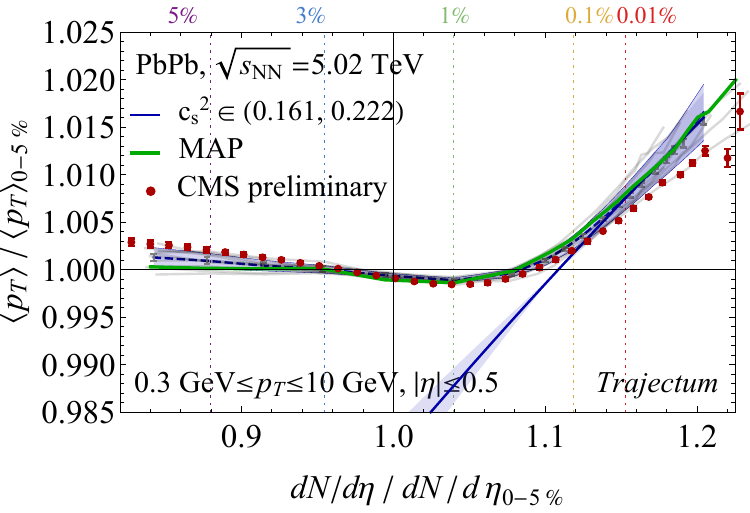}
\includegraphics[width=0.45\textwidth]{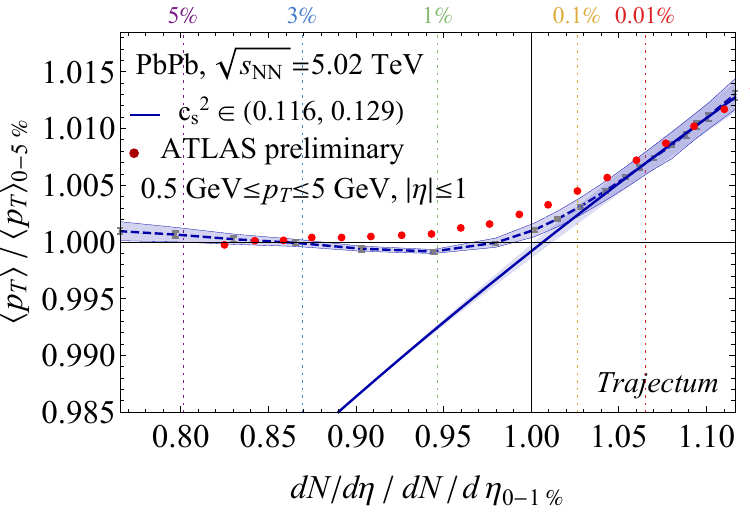}
\caption{\label{fig:cmsmainresult03}(left) We show the equivalent figure as Fig.~\ref{fig:cmsmainresult}, but now only taking charged particles with $p_T>0.3\,$GeV\@. This has the advantage that experimentally no extrapolation to $p_T > 0$ is required, and hence a full theory to experiment comparison can be made. Perhaps surprisingly, also the theoretical systematic uncertainty is significantly smaller. (right) We show the equivalent figure as Fig.~\ref{fig:cmsmainresult} for our MAP settings, but now only taking the same cuts and centrality selection as the $N_{\rm ch}^{\rm rec}$ method of \cite{ATLAS:2023xpwmanual}\@. Similar to Fig.~\ref{fig:cmsmainresult} we find a good agreement, albeit with a slightly higher slope than found in experiment.}
\end{figure*}

In Fig.~\ref{fig:cmsmainresult03} (left) we compare the $p_T > 0.3\,$GeV setting with the CMS preliminary result \cite{CMS:2023byutalk}\@. Consistent with the theoretical predictions CMS also finds a smaller slope than Fig.~\ref{fig:cmsmainresult}\@. The CMS uncertainties are much smaller, since no (unknown) extrapolation to $p_T > 0$ is necessary, but it is noteworthy that also the theoretical systematic uncertainties are significantly smaller.

Lastly, in Fig.~\ref{fig:cmsmainresult03} (right) we compare to the similar results from the ATLAS collaboration \cite{ATLAS:2023xpwmanual}, which use a $0.5\,$GeV $p_T$ cut as well as a multiplicity based centrality selection using the same cut at mid-rapidity. As in Fig.~\ref{fig:cmsmainresult} we find that \tr is slightly above the experimental result.

\subsection{Predictors}

\begin{figure*}
\includegraphics[width=0.85\textwidth]{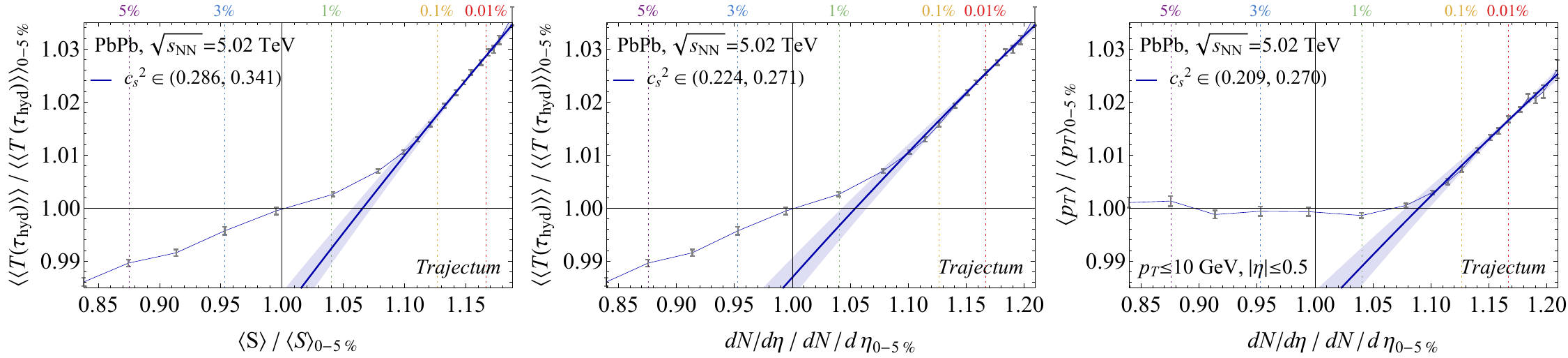}
\caption{\label{fig:predictors}We show temperature versus total entropy (left), temperature versus multiplicity (middle) and \mpt{} versus multiplicity (right)\@. If multiplicity is proportional to the total entropy and the mean tranverse momentum to the temperature then all these curves should be equivalent. We see that the slope in particular decreases when going from entropy to multiplicity, which at least partly can be explained by Poisson fluctuations. For larger centralities around 5\% we also see a non-trivial slope in the average temperature that is not present for the \mpt{}\@. In that regime we hence do not exactly satisfy the linear relation between \mpt{} and temperature as proposed in \cite{Gardim:2019brr}.
}
\end{figure*}

In \cite{Gardim:2019brr} relatively simple arguments were used to obtain Eqn.~\eqref{eq:ptcs2}, which includes assuming that multiplicity is proportional to the total initial entropy and that the \mpt{} is proportional to the initial average temperature. As such it is interesting to verify whether these arguments are consistent with the full \tr simulations as done in this work. Some insight can be obtained by looking at Fig.~\ref{fig:predictors}, which shows temperature versus entropy (left), temperature versus multiplicity (middle) and finally transverse momentum versus multiplicity (right), all for the same MAP settings. 

The second and third slope are consistent, which could indicate that the temperature may indeed be a good estimator for the transverse momentum, at least for our 'standard' centrality selection (see Fig.~\ref{fig:varycentrselection} in the main text)\@. Switching from entropy to multiplicity decreases the slope and increases the fluctuations (note the x-axis shift), which at least partly can be attributed to Poisson fluctuations (reference \cite{Gardim:2019brr} estimates an effect of 10\% for the ATLAS acceptance, but this will be larger in our simulation since we use a rapidity range of only $|\eta|<1.5$).

We note that to then relate the slope to the speed of sound one also needs a constant initial size of the plasma, which is not entirely trivial (see Fig.~\ref{fig:geometryucratio}).

\end{document}